\documentclass[pra,superscriptaddress,twocolumn]{revtex4-1}

\usepackage{amsfonts,amssymb,amsmath}
\usepackage[]{graphics,graphicx,epsfig}
\usepackage{amsthm}

\def\identity{\leavevmode\hbox{\small1\kern-3.8pt\normalsize1}}

\newcommand{\tr}[0]{{\rm Tr}}
\def\openone{\leavevmode\hbox{\small1 \normalsize \kern-.64em1}}

\def\tr{{\rm Tr}}

\def\<{\langle}
\def\>{\rangle}

\newtheorem{propo}{Proposition}

\newcommand{\be}{\begin{eqnarray}}
\newcommand{\ee}{\end{eqnarray} }
\newcommand{\bpr}{\begin{propo}}
\newcommand{\epr}{\end{propo}}
\newcommand{\bpf}{\begin{proof}}
\newcommand{\epf}{\end{proof}}

\newcommand{\Tr}{\mathrm{Tr}}

\renewcommand{\epsilon}{\varepsilon}

\begin{document}

\title{Unified approach to contextuality, non-locality, and temporal correlations}

\author{M. Markiewicz}
\affiliation{Centre for Quantum Technologies, National University of Singapore, Singapore}
\affiliation{Institute of Theoretical Physics and Astrophysics, University of Gda\'nsk, Poland}

\author{P. Kurzy\'nski}
\affiliation{Centre for Quantum Technologies, National University of Singapore, Singapore}
\affiliation{Faculty of Physics, Adam Mickiewicz University, Poland}

\author{J. Thompson}
\affiliation{Centre for Quantum Technologies, National University of Singapore, Singapore}

\author{S.-Y. Lee}
\affiliation{Centre for Quantum Technologies, National University of Singapore, Singapore}

\author{A. Soeda}
\affiliation{Centre for Quantum Technologies, National University of Singapore, Singapore}

\author{T. Paterek}
\affiliation{Centre for Quantum Technologies, National University of Singapore, Singapore}
\affiliation{School of Physical and Mathematical Sciences, Nanyang Technological University, Singapore}

\author{D. Kaszlikowski}
\email{phykd@nus.edu.sg}
\affiliation{Centre for Quantum Technologies, National University of Singapore, Singapore}
\affiliation{Department of Physics, National University of Singapore, Singapore}

\begin{abstract}
We highlight the existence of a joint probability distribution as the common underpinning assumption behind Bell-type, contextuality, and Leggett-Garg-type tests.
We then present a procedure to translate contextual scenarios into temporal Leggett-Garg-type and spatial Bell-type ones.
To demonstrate the generality of this approach we construct a family of spatial Bell-type inequalities.
We show that in Leggett-Garg scenario a necessary condition for contextuality in time is given by a  violation of consistency conditions in Consistent Histories approach to quantum mechanics.
\end{abstract}

\maketitle

\section{Introduction}
Classical physical theories such as general relativity, electrodynamics and thermodynamics describe a universe where acts of observation merely reveal underlying reality.
For instance, an electromagnetic field or a black hole exists objectively at all times independently of whether we choose to observe it or not.
Quantum theory of matter is different.
It was theoretically demonstrated by Bell in 1964~\cite{Physics.1.195} and by Kochen and Specker (KS) in 1967~\cite{JMathMech.17.59} that quantum mechanical predictions depend on the act of observation.
Both of these predictions found confirmation in experiments:
Bell inequalities were violated in numerous laboratories~\cite{PhysRevLett.47.460,PhysRevLett.49.1804,PhysRevLett.81.5039,Nature.409.791,ProcNatAcadSci.107.19708}
as were certain inequalities encapsulating KS ideas~\cite{Nature.474.490,arXiv.1301.2887}.
In 1985 Leggett and Garg (LG) presented a related notion of macroscopic realism~\cite{PhysRevLett.54.857}
asserting that a macroscopic system should at all times be in one of its macroscopically distinguishable states that do not change if a measurement is performed on the system.

Although the theorems by Bell, KS and LG seem different they are in fact based on the same underlying hypothesis.
They all contrast quantum mechanical predictions with predictions of theories that assume the existence of a joint probability distribution for the outcomes of all possible measurements one can perform on a physical system.
More precisely, consider measurements from the set $S=\{X_1,X_2,\dots,X_N\}$, such that measurement $X_j$ yields an outcomes $x_j$, with $j=1,2,\dots, N$.
For some of these measurements a joint probability distribution, of the type $p_\mathrm{exp}(x_i,x_j)$, can be experimentally obtained
while for other subsets such experimental joint probability distributions cannot be measured.
For example, according to quantum theory it is impossible to construct a device capable of simultaneously measuring two non-commuting observables on a single system.
Objective reality assumes that nevertheless there exists a joint probability distribution for the full set of these observables, $p(x_{1},x_{2},\dots,x_{N})$.
Depending on a physical scenario considered, the lack of such joint probability is called quantum non-locality, contextuality or violation of macroscopic realism.

In Bell-type experiments the set $S$ is divided into two or more groups of measurements $S = A \cup B \cup \dots$, such that each group represents a set of measurements performed by spatially separated observers.
If one arranges a situation in which the measurements are space-like separated, special relativity dictates a natural assumption that the outcomes obtained on individual systems do not depend on the parameters in distant laboratories.
In such a case the existence of the joint probability distribution is known as the assumption of local realism~\cite{PhysRevLett.48.291}, first formulated in 1935 in the important paper by Einstein, Podolsky and Rosen~\cite{PhysRev.47.777}.
The fact that quantum mechanical predictions cannot be described in this way is sometimes phrased as quantum non-locality.

In KS-type experiments there are no spatially separated systems.
For instance, the simplest KS scenario introduced by Klyachko \emph{et. al.}~\cite{PhysRevLett.101.20403} contains a set of five measurements for which one can experimentally establish joint probabilities $p_\mathrm{exp}(x_1,x_2), p_\mathrm{exp}(x_2,x_3), \dots, p_\mathrm{exp}(x_5,x_1)$.
If the joint probability distribution for the outcomes of all these observables exists, $p(x_1,x_2,\dots,x_5)$, such a model is known as a non-contextual realistic theory.
In this sense, each Bell-type experiment is a special case of KS experiment where the context of measurements is provided by spatial separation of observers.

LG-type scenarios are similar to the KS-type experiments in that a single physical system is being interrogated.
In the LG setting there is a single physical property $X_t$ that evolves in time.
This property is measured at different times $t_1,t_2,\dots$ and probabilities $p_\mathrm{exp}(x_{t_i},x_{t_j})$ are estimated for suitable time slices.
The existence of the joint probability distribution for the outcomes at all times, $p(x_{t_1},x_{t_2},\dots)$, whose marginals agree with $p_\mathrm{exp}(x_{t_i},x_{t_j})$, is known as the assumption of macroscopic realism.
In quantum mechanics the lack of this joint probability distribution is due to intermediate quantum evolution and the invasive nature of quantum measurements.

Since all these cases share the same mathematical background one expects to find correspondence between them.
Here we reveal this correspondence and use it to derive new inequalities.
First we review a simple test of contextuality of a single system and then explain how it can be extended to both a temporal scenario and a spatial scenario on two subsystems.
Next, we discuss how to translate between general correlations of two measurements realised in three different scenarios: contextual, non-local and temporal.
We conclude with observations relating our correspondence to consistent histories and quantum cryptography.

\section{Compatible measurements on a single system}
We first study the scheme proposed by Klyachko-Can-Binicioglu-Shumovsky (KCBS)~\cite{PhysRevLett.101.20403}.
Consider five dichotomic $\pm 1$ measurements $X_j$ on a single system where each $X_j$ is compatible with $X_{j-1}$ and $X_{j+1}$ for $j=0,\ldots,4$, and sums are modulo $5$.
Compatible, here, means from the operational point of view that these observables can be measured jointly or sequentially, with an assumption that sequential measurements do not affect each other.
More precisely, if $X_j$ is measured first, then the measurement of $X_i$ does not change the outcome of $X_j$, which can be confirmed by a subsequent measurement of $X_j$, i.e., the measurement sequence is $X_j \rightarrow X_i \rightarrow X_j$.  This property guarantees non-invasiveness of measurements.

The possibility of joint probability distribution of the outcomes of all physical observables on a single system can be tested by the following KCBS inequality proposed by Ref. \cite{PhysRevLett.101.20403}:
\be
\sum_{j=0}^4 \langle X_j X_{j+1} \rangle  \ge -3.
\ee
For completeness we present a proof of this inequality.
By definition each correlation function is given by
\be
\<X_iX_j\>=\sum_{x_i,x_j=\pm1}x_ix_j p(x_i,x_j).
\ee
By assumption there exists a joint probability distribution for all variables $X_i$, e.g.:
\be
&&\<X_0X_1\>=\sum_{x_0,...,x_4=\pm1} x_0x_1 p(x_0,x_1,x_2,x_3,x_4).
\ee
Note that for a non-contextual assignment of values $x_j = \pm 1$ we have $\sum_{j=0}^4 x_j x_{j+1} \geq -3$, which can be directly verified.
Combining all above expressions we get:
\be
\label{kcbs}
& &\<X_0X_1\>+\<X_1X_2\>+\<X_2X_3\>+\<X_3X_4\>+\<X_4X_0\>\nonumber\\
& = & \sum_{x_0,...,x_4=\pm1}p(x_0,x_1,x_2,x_3,x_4) \sum_{j=0}^4 x_j x_{j+1} \nonumber \\
& \geq & \sum_{x_0,...,x_4=\pm1}p(x_0,x_1,x_2,x_3,x_4) (-3) = -3. \nonumber
\ee

In quantum mechanics the compatibility is provided by $[X_j, X_{j \pm 1}] = 0$.
Maximal quantum violation of the above inequality (so-called Tsirelson bound) within this framework is known to be \cite{CSW10,GBCKL12}:
\be
T_{\mathrm{context}}=5-4\sqrt{5}\approx-3.94.
\ee

\section{Temporal KCBS inequality}
Instead of studying contextuality  using the KCBS inequality we investigate a temporal-non-contextual-inequality whose construction roughly parallels seminal work done by Leggett and Garg \cite{PhysRevLett.54.857} and continued in \cite{B09, AHW10}.

Consider a dichotomic $\pm 1$ measurement, $X_t$, which is conducted at time $t = \{t_0,t_1,\dots,t_4\}$.
If we make successive measurements at two sequential times, then we can construct two point temporal correlations
$\< X_{t_0} X_{t_{1}}\>, \< X_{t_1} X_{t_{2}}\>, \< X_{t_2} X_{t_{3}}\>, \< X_{t_3} X_{t_{4}}\>, \< X_{t_0} X_{t_{4}}\>$.
These two-point temporal correlations naturally lead to a  temporal analogue of the KCBS inequality
\be
\label{tempkcbs}
&& \< X_{t_0} X_{t_{1}}\> + \< X_{t_1} X_{t_{2}}\> + \< X_{t_2} X_{t_{3}}\> \nonumber \\
&+ & \< X_{t_3} X_{t_{4}}\> + \< X_{t_0} X_{t_{4}}\> \geq - 3.
\ee
This inequality will be satisfied whenever there is a joint probability distribution which ascribes predetermined outcomes to the measurements $X_t$ at all times $t_0,\dots, t_4$.

The existence of the joint probability distribution in this scenario is tantamount to Leggett and Garg's ``macrorealism" condition~\cite{PhysRevLett.54.857}.
Conversely, violation of the inequality (\ref{tempkcbs}) can be called  \emph{contextuality in time}.

In quantum mechanics the inequality (\ref{tempkcbs}) can be violated using a single spin-$\frac{1}{2}$ particle.
We stipulate that in each run of the experiment we make precisely two measurements corresponding
to a pair of observables $X_{t_i}$ and $X_{t_{i \pm 1}}$.
For definitiveness, we specify the observable $X_{t}$ to be represented by a $\sigma_z$ Pauli operator measured at one of five distinct times, $t \in \{t_0,\dots, t_4 \}$.
We initialise the spin in a completely mixed state and allow it to evolve under the unitary operator
\be
U=e^{i \frac{8}{5}\pi t\sigma_y}.
\ee
For this scenario, the left-hand side of the inequality (\ref{tempkcbs}) attains the minimal value of $\approx -4.045$ if we choose the time instances $t\in\{t_0,\dots, t_4\} = \{0,\frac{1}{4},\frac{1}{2},\frac{3}{4},1\}$.
It has recently been proved by G\"uhne \emph{et. al.} \cite{GBCKL12} that this is the maximum possible violation of inequality (\ref{tempkcbs}) by a qubit when pairs of sequential measurements do not commute.
In order to calculate temporal correlations of $\pm 1$ measurements we use the following formula
\be
\<X_{t_1} X_{t_2}\> & = & p_{+1} q_{+1|+1} + p_{-1}q_{-1|-1}\nonumber\\
&-& p_{+1}q_{-1|+1} - p_{-1}q_{+1|-1},
\ee
where $p_{k}$ denotes probability of outcome $k$ in the first measurement (at instant $t_1$), and $q_{l|k}$ denotes probability of the outcome $l$ in the second measurement (at $t_2$) on condition that outcome $k$ occured in the first one.
In quantum mechanics this formula reduces to~\cite{F10}
\be
\label{anti}
\<X_{t_1} X_{t_2} \>=\frac{1}{2}\tr\left(\rho\{X_{t_1},X_{t_2}\}\right),
\ee
where $\{X_{t_1},X_{t_2}\}$ denotes the anti-commutator.

\section{New spatial inequality and its quantum violation}
Finally, the inequality (\ref{tempkcbs}) can be transformed into a Bell-type inequality testing the existence of a joint probability distribution for spatially separated local measurements.
Within this framework $\< X_i X_j \> = \< A_i B_j \>$ are correlations obtained on space-like separated systems $A$ and $B$.
The inequality (\ref{tempkcbs}) takes the form:
\be
\label{kcbssp}
&&\<A_0 B_1\>+\< A_1 B_2\>+\<A_2 B_3\>+\< A_3B_4\>+\<A_4 B_0\>\geq-3\nonumber\\
\ee
with additional constraint
\be
\label{kcbsbell}
\< A_i B_i\>=1 \quad \textrm{ for all } i,
\ee
which means that $A_i$ and $B_i$ always have the same outcomes.

The inequality (\ref{kcbssp}) together with the assumption $\langle A_i B_i \rangle=1$ resembles the original Bell scenario for three $\pm 1$ qubit measurements $A$, $B$ and $C$ \cite{Physics.1.195}:
\be
1+\langle B \otimes C\rangle\geq |\langle A \otimes B\rangle + \langle A \otimes C\rangle|,
\ee
where it was assumed that $\langle B\otimes B\rangle = -1$ due to correlations of the singlet state. The additional assumption of the outcome correlation of pairs of spatially separated measurements is often considered as a weakness of this type of nonlocality tests. In real experimental scenarios other tests that do not require this assumption are preferred. Nevertheless, the inequality (\ref{kcbssp}) can be used as a theoretical tool to refute local realistic description of quantum measurements and, what is more important, to establish a unified framework to describe contextuality, nonlocality and contextuality in time as different physical manifestations of the violation of the same mathematical property.

The optimal violation of (\ref{kcbssp}) can be obtained for the state $\rho'=|\phi_+\rangle\langle\phi_+|$, with $|\phi_+\rangle=\frac{1}{\sqrt{2}}(|00\rangle+|11\rangle)$, and for measurements $A_i = \sigma_i\otimes\openone$, $B_i= \openone\otimes\sigma_i$, where $i=0,\dots,4$ and $\sigma_i= e^{i\frac{2\pi i}{5}\sigma_y}\sigma_z e^{-i\frac{2\pi i}{5}\sigma_y}$.
Note that the state $|\phi_+\rangle$ has the property that for the qubit measurements in the $xz$-plane, $M(\alpha)=\cos\alpha \, \sigma_z + \sin\alpha \, \sigma_x$, one has $\langle \phi_+ | M(\alpha)\otimes M(\alpha)|\phi_+\rangle=1$. Therefore, the assumption $\langle A_i B_i \rangle=1$ is fulfilled by this state.

On the level of quantum mechanics the link between temporal and spatial correlations was noticed before~\cite{F10}.
This is a direct consequence of an extension of  Tsirelson's theorem on different representations of correlation matrices for quantum observables \cite{T87}.
It says that the following two statements are equivalent:
\begin{enumerate}
        \item There exists a Hilbert space $\mathcal H$ together with Hermitian operators $A_1,\ldots,A_m,B_1,\ldots,B_n\in \mathcal B(\mathcal H)$ fulfilling $A_k^2=\openone,\,B_l^2=\openone$, and a density matrix $\rho$ such that:
\be
\<A_kB_l\>=\frac{1}{2}\tr(\rho\{A_k,B_l\}).
\ee
        \item There exist Hilbert spaces $\mathcal H_A$ and $\mathcal H_B$ together with Hermitian operators $A_1,\ldots,A_m\in \mathcal B(\mathcal H_A)$, $B_1,\ldots,B_n\in \mathcal B(\mathcal H_B)$ fulfilling $A_k^2=\openone,\,B_l^2=\openone,\nonumber$, and a density matrix $\rho'$ on $\mathcal H_A\otimes \mathcal H_B$ such that:
\be
\<A_kB_l\>=\tr(\rho'(A_k\otimes B_l)).
\ee
\end{enumerate}
What is very important and omitted in \cite{F10} is that
the state $\rho'$, due to Tsirelson's construction, has a very specific form.
Namely, it has to fulfil the following relations:
\be
\<A_iB_i\>&=&\tr(\rho' A_i\otimes B_i)=1,\,\,\textrm{ for }i=1,\ldots,\min(m,n).\nonumber
\ee
Equivalence between the two above statements implies that the Tsirelson bound of the Bell-type inequality (\ref{kcbssp}) is the same as for the temporal  inequality (\ref{tempkcbs}).

\section{Generalization to arbitrary number of measurements}
The approach to the KCBS scenario discussed above can be generalised in a straightforward manner to any test of contextuality that utilises two-point correlations.
Up to now we discussed scenarios involving five measurements.
In general, if one can experimentally evaluate $p_\mathrm{exp}(x_i,x_{i+1})$ for $N$ dichotomic $\pm 1$ measurements $i=0,\dots,N-1$,
the existence of joint probability distribution $p(x_0,\dots, x_{N-1})$ implies that the following inequality is satisfied~\cite{AQBT,GBCKL12}:
\be
\sum_{i=0}^{N-2} \< X_i X_{i+1}\> + (-1)^{N-1} \< X_{N-1} X_{0}\> \ge -N+2.
\label{BR}
\ee
Using our framework this inequality can be tested in three physical settings.

The first one is the contextuality test on a single system where co-measurability is provided by compatibility of measurements,
i.e. there exists a device for which the outcomes of both measurements are always independent of the order in which they are performed.

In the second, temporal, setting one can treat the two-point correlations entering (\ref{BR}) as expectation values of two measurements performed at different times.
The measurements here are no longer required to be compatible and the co-measurability is provided by temporal separation.

The implementation of the third, spatial, scenario depends on the parity of $N$.
If $N$ is an even number there exist a natural bipartition of measurements, $X_{2i} = A_i$ and $X_{2i+1} = B_i$, and the inequality is transformed to:
\begin{multline}
\label{Neven}
\< A_0 B_{0}\> + \< A_1 B_{0}\> + \< A_1 B_{1}\> + \dots  + \< A_{(N-2)/2} B_{(N-2)/2}\> \\
- \< A_0 B_{(N-2)/2}\>  \ge -N+2.
\end{multline}
However, in the case of odd $N$, such bipartition does not exist and   to bypass this problem we propose  to double the number of measurements, i.e., Alice (Bob) has $N$ measurements $A_0,\dots,A_{N-1}$ ($B_0,\dots,B_{N-1}$).
In addition we require perfect correlations between the corresponding local measurements, i.e. $\< A_i B_i \> = 1$ for all $i$.
This implies that the outcome of observable $A_i$ and $B_i$ is always the same.
Consider now the following inequality
\be
\sum_{i=0}^{N-2} \< A_i B_{i+1}\> + (-1)^{N-1} \< A_{N-1} B_{0}\> \ge -N+2.
\label{BR_SPATIAL}
\ee
The local realistic bound of $-N+2$ follows from the fact that although $\< A_i A_{i+1}\>$ is not directly measurable,
due to our assumptions $\< A_i A_{i+1}\> = \< A_i B_{i+1}\>$ and one can rewrite this spatial inequality using only the measurements of Alice in which case the inequality has the form of (\ref{BR}).

The  Bell inequalities \eqref{Neven} and \eqref{BR_SPATIAL} are violated by quantum measurements on a $|\phi^+\rangle$ state for arbitrary $N$. In case of (\ref{Neven}) Alice measures $N/2$ observables $A_m$, given by corresponding Bloch vectors:
$$\vec a_m = (-\sin(2m\pi/N), 0, \cos(2m\pi/N)),$$
whereas Bob's observables $B_m$ are given  by Bloch vectors:
$$\vec b_m = \left(\sin((2m+1))\pi/N), 0, -\cos((2m+1))\pi/N\right).$$
In case of odd $N$ Alice and Bob measure $N$ observables $A_m, B_m$ given by Bloch vectors:
$$\vec a_m=\vec b_m=(\sin((\pi-\pi/N)m), 0, \cos((\pi-\pi/N)m)).$$
These settings give rise to violation for any $N$:
\begin{equation}
N \cos(\pi(N-1)/N) < - N + 2
\end{equation}

We also note that although the bounds resulting from the existence of a joint probability distribution are by construction the same in all three scenarios, the quantum Tsirelson bounds may be different.
For example, in the case of KCBS scenario we showed that the quantum contextual bound differs from the quantum temporal and non-local bounds.
This is due to the difference in the compatibility criterion in the respective scenarios.
It would be interesting to investigate further if this difference is present for all tests involving an odd number of measurements~\cite{GBCKL12}.

\section{Consistent Histories approach to temporal inequalities}
The violation of local realism and non-contextuality is a result of non-classical correlations between subsystems or between local observables. In quantum theory these correlations stem from entanglement or from commutation properties of local operators that are used to test non-contextuality. On the other hand, non-classical temporal correlations result from the lack of commutativity between the observables (in Heisenberg picture) that are sequentially measured. We now show that these correlations can be also interpreted as resulting from inconsistency of histories describing the measurement scenario. This is done within the Consistent Histories approach to quantum theory \cite{CH, O92}.
The result might be anticipated as satisfying the consistency conditions allows ordinary probabilistic reasoning~\cite{CH,O92} and guarantees non-invasiveness of intermediate measurements~\cite{DESPAGNAT}.
Nevertheless, it is instructive to see how the consistency condition naturally emerges in the context of temporal inequalities.

Consider two sequences of events $(e_1,e_2,\dots,e_N)$ and $(f_1,f_2,\dots,f_N)$. We assume that these events are ordered in time, i.e., $e_i$ happens before $e_j$ if $i<j$ (similar for $f_i$). We refer to these sets as  'history $e$' and 'history $f$'. In quantum theory these events correspond to projectors $P^e_1,P^e_2,\dots,P^e_N$ and $P^f_1,P^f_2,\dots,P^f_N$. Next, consider operators $C_e=P^e_N\dots P^e_2 P^e_1$ and $C_f=P^f_N\dots P^f_2 P^f_1$.
It is said that the two histories measured on a state $\rho$ are consistent if and only if \cite{O92}:
\begin{equation}\label{ch0}
\text{Re}\left(\text{Tr}\left(C_e\rho C_f^{\dagger}\right)\right)=0.
\end{equation}
This condition assures validity of ordinary probabilistic reasoning about joint events without going to any contradictions.

Next, consider the LG inequality in the form
\begin{equation}\label{ch1}
\langle X_1 X_2\rangle + \langle X_2 X_3\rangle + \langle X_1 X_3\rangle \geq -1,
\end{equation}
where $X_i$ ($i=1,2,3$) are $\pm 1$ observables. We associate the corresponding measurement events with projectors $P^{(i)}_{k}$, where $k=\pm 1$. Let us define a probability distribution for all three measurement outcomes $p(X_1=k,X_2=l,X_3=m) \equiv p(k,l,m)$ as
\begin{equation}\label{ch2}
p(k,l,m)=\text{Tr}\left(P^{(3)}_{m}P^{(2)}_{l}P^{(1)}_{k}\rho P^{(1)}_{k}P^{(2)}_{l}P^{(3)}_{m}\right),
\end{equation}
where the last term $P^{(3)}_{m}$ is not necessary, but we put it for convenience. Note, that this probability distribution does not necessarily reproduce marginal probabilities, therefore it may not be a joint probability distribution that guarantees a classical model. It rather provides us with a link to the consistent histories formalism, where $p(k,l,m)$ are probabilities of histories $(k,l,m)$.
Each quantum correlation function entering (\ref{ch1}) can be expressed in analogy to
\begin{equation}\label{ch3}
\langle X_2 X_3\rangle=\sum_{k} \left(p(*,k,k)-p(*,k,-k)\right),
\end{equation}
where, e.g., $p(*,k,k) = \Tr(P^{(3)}_{k}P^{(2)}_{k}\rho P^{(2)}_{k}P^{(3)}_{k})$ is calculated by considering a hypothetical measurement at time $t_1$, and using $\sum_k P^{(1)}_k=\openone$:
\begin{equation}
p(*,k,k) = p(+,k,k) + p(-,k,k) + I(*,k,k),
\label{CH_MARG_I}
\end{equation}
where we introduced the interference term $I(*,k,k) = 2 \text{Re}\left(\text{Tr}\left(P^{(3)}_{k}P^{(2)}_{k}P^{(1)}_{+}\rho P^{(1)}_{-}P^{(2)}_{k}P^{(3)}_{k}\right)\right)$.
We therefore arrive at the following form of inequality (\ref{ch1}):
\begin{eqnarray}\label{eq:contextineq}
\sum_k \left(4p(k,k,k) + I(*,k,k) + I(k,*,k) \right. \nonumber \\ \left. - I(*,k,-k) - I(k,*,-k)\right) \geq 0, \label{ch4}
\end{eqnarray}
where we also used the fact that $I(k,k,*) = I(k,-k,*) = 0$.
A necessary condition for the violation of the inequality (\ref{ch4}) is that at least one of the terms $I(*,k,k), I(k,*,k), I(*,k,-k), I(k,*,-k)$ is nonzero. This however implies that at least one pair of the histories:
\be
\label{allhist}
& &\{(+,k,k),(-,k,k)\},\{(k,+,k),(k,-,k)\}\nonumber\\
& &\{(+,k,-k),(-,k,-k)\},\{(k,+,-k),(k,-,-k)\},
\ee 
contains inconsistent ones, which follows directly from (\ref{ch0}). 

A direct validation of this statement that does not utilise inequalities
comes from noticing that $p(k,l,m)$ provides a valid marginal probability in (\ref{CH_MARG_I}) only if $I(*,k,k) = 0$, ergo the histories $(+,k,k)$ and $(-,k,k)$ are consistent.

\begin{figure}
\includegraphics[width=0.45\textwidth]{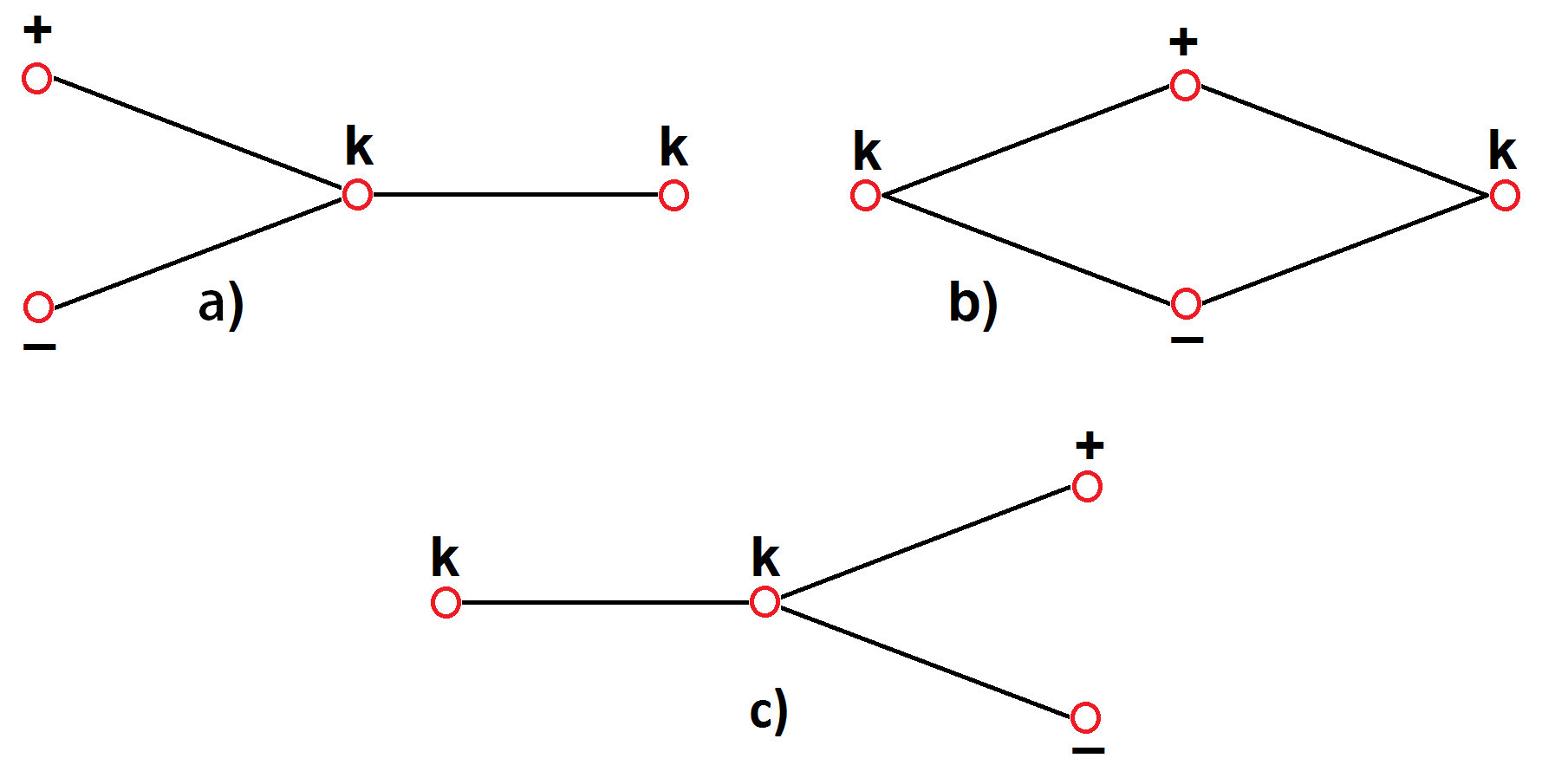}
\caption{Pictorial description of the histories related to interference terms in Eq.\eqref{eq:contextineq}; a) histories $\{(+,k,k),(-,k,k)\}$, related to the term $\langle X_2 X_3\rangle$, can give rise to nonzero interference term $I(*,k,k)$; b) histories $\{(k,+,k),(k,-,k)\}$, related to the term $\langle X_1 X_3\rangle$, can give rise to nonzero interference term $I(k,*,k)$; c) histories $\{(k,k,+),(k,k,-)\}$, related to the term $\langle X_1 X_2\rangle$ are always consistent, therefore there is no interference of the form $I(k,k,*)$.
}
\label{fig1}
\end{figure}

It is worth mentioning that the vanishing interference terms $I(k,k,*)$ and $I(k,-k,*)$ are related to the term $\langle X_1 X_2\rangle$. Therefore only the terms $\langle X_1 X_3\rangle$ and $\langle X_2 X_3\rangle$ directly give rise to the quantum violation of the LG inequality \eqref{ch1} (Fig. \ref{fig1}).
\\

Intuitively, the consistency conditions assure that the additivity of probabilities of single events is compatible with additivity of squared quantum probability amplitudes \cite{O92}. On the contrary, violation of these conditions implies that some interference terms between the probability amplitudes arise.


\section{Conclusions}
We discussed Bell, Kochen-Specker, and Leggett-Garg experiments and showed that they are all different physical manifestations of the violation of the same underlying mathematical property --- the existence of a joint probability distribution for all possible measurements that can be performed on the system.
We introduced correspondence between these scenarios.

Note that this correspondence can be used to establish a link between the two acclaimed quantum key distribution protocols,
the BB84~\cite{BB84} protocol and the Ekert protocol~\cite{PhysRevLett.67.661}.
Although the security of both protocols relies on different fundamental physical principles,
mathematically speaking their security stems from the lack of a joint probability distribution.
The Ekert protocol utilises quantum non-locality whereas BB84 relies on invasiveness of quantum measurements,
effectively contradicting the assumptions of the macro-realism.

Utilizing the Consistent Histories approach to sequential measurements we found, that a necessary condition for violation of temporal Bell inequalities is existence of interference effects between probability amplitudes related to sequences of events.

We are confident this general framework will find further applications.
For instance, one can easily approach the problem of mixed space and time quantum correlations \cite{DASH13}
and has an attractive feature that it can be implemented numerically using standard modules for linear programming.

\section*{Acknowledgments}
This work is supported by the National Research Foundation and Ministry of Education in Singapore.
PK is supported by the Foundation for Polish Science.
TP acknowledges NTU start-up grant.
The contribution of MM is supported within the International PhD Project
``Physics of future quantum-based information technologies''
grant MPD/2009-3/4 from Foundation for Polish Science and by the NCN Grant No. 2012/05/E/ST2/02352.

During the completion of this work two independent papers covering similar topics appeared~\cite{GBCKL12, DASH13}.

\end{document}